**"I would love this to be like an assistant, not the teacher": a voice of the customer perspective of what distance learning students want from an Artificial Intelligence Digital Assistant**

Bart Rienties[1], John Domingue[2], Subby Duttaroy[3], Christothea Herodotou[1], Felipe Tessarolo[1], Denise Whitelock[1].

[1] The Institute of Educational Technology, The Open University.

[2] Knowledge Media Institute, The Open University

[3] Business Development Unit, The Open University.

**Abstract**

With the release of Generative AI systems such as ChatGPT an increasing interest in using Artificial Intelligence (AI) has been observed across domains, including higher education. While emerging statistics show the popularity of using AI amongst undergraduate students, little is yet known about students' perceptions regarding AI including self-reported benefits and concerns from their actual usage, in particular in distance learning contexts. Using a two-step, mixed-methods approach, we examined the perceptions of ten online and distance learning students from diverse disciplines regarding the design of a hypothetical AI Digital Assistant (AIDA). In the first step, we captured students' perceptions via interviews, while the second step supported the triangulation of data by enabling students to share, compare, and contrast perceptions with those of peers. All participants agreed on the usefulness of such an AI tool while studying and reported benefits from using it for real-time assistance and query resolution, support for academic tasks, personalisation and accessibility, together with emotional and social support. Students concerns related to the ethical and social implications of implementing AIDA, data privacy and data use, operational challenges, academic integrity and misuse, and the future of education. Implications for the design of AI-tailored systems are also discussed.

# Introduction

Artificial Intelligence (AI) has been around for over half a century. However, with the rapid evolution of AI applications in the last five years, in particular in the field of so-called Generative AI [Gen AI:1, 2, 3] and popular applications like ChatGPT and Bard (now branded as Gemini), advocates of technology [4, 5], policymakers [6, 7], scientists [1, 3, 8, 9], and tech companies like Google, Meta, and OpenAI argue that we are at a dawn of a new digital and societal revolution.

For many decades Artificial Intelligence in Education (AIED) researchers have been chasing the "Bloom dream" [10], whereby AI technology can dramatically uplift student attainment through the provision of high-quality personalised learning. The arrival of ChatGPT in November 2022, exposing the affordances of AI based on Large Language Models (LLM) outside of dedicated research laboratories, has raised the prospects of this dream becoming a reality [1, 3, 11, 12]. With over 1.6 billion visits of ChatGPT in December 2023 in multiple languages, and a reported weekly user base of 100 million users, ChatGPT is most likely being used by many students as well as educators [13, 14]. Emerging statistics of student usage of AI tools in November 2023 show that more than half of UK students (53%) consult AI for marked work, with one in four using ChatGPT or BARD to identify topics [14].

In a recent tertiary (meta) review of 66 systematic literature reviews of AI in formal higher education or continuing education settings, Bond, Khosravi, De Laat, Bergdahl, Negrea, Oxley, Pham, Chong and Siemens [8] indicated that the most reported positive benefit of the use of AI in education was personalised learning (38.7%), followed by greater insights into student understanding, positive influence on learning outcomes, and reduced planning and administration for educators. At the same time, in terms of the most evident research gaps identified by [8] across these studies were the ethical implications of using AIED (40.9%), the lack of diversity in methodological approaches (36.4%), specific needs to apply research in wider educational practice (33.3%), and with a wider range of stakeholders beyond STEM (21.2%).

There is a paucity of research on how distance learning students think about the affordances and limitations of AI for their studies. In particular, given the increased focus on online and distance education, it is surprising that most of the research focussed on Gen AI applications like ChatGPT seem to be focussed on undergraduate STEM and Computer Science students studying at on-campus universities [8]. In line with recommendations by [8] for more research using "qualitative, mixed methods and design-based approaches" and include "students perceptions of the effectiveness and AI fairness", in particular from underrepresented groups and students with disabilities, we developed a design-based mixed method study.

In this first research cycle we involved students from a range of disciplines at the early stages of the development of so-called AI Digital Assistant (AIDA). Using a so-called voice of the customer approach [VoC: 15], in a two-stage sequential approach of interviews with a subsequent online survey, we aim to explore what distance learners expect in terms of services of AIDA for their studies, and what their concerns might be.

## Supporting distance learners with AI

Over the years a range of AI applications have been developed that might help to support the Bloom dream. For example, in a review by Zawacki-Richter, Marín, Bond and Gouverneur [16] of 146 AIED articles four common AIED typologies were identified: 1) profiling and prediction [17-20]; 3) intelligent tutoring systems [12, 21, 22]; 3) assessment & evaluation [23-26]; 4) adaptive systems and personalisation [27, 28].

Practical applications of such AIED systems include automatic grading [24, 25], digital avatars [29], chatbots [30], intelligent tutoring systems [12], recommending personalised content [31], and robots [32]. These four common AIED typologies were also used in the subsequent meta-review of Bond, Khosravi, De Laat, Bergdahl, Negrea, Oxley, Pham, Chong and Siemens [8] in order to classify which AIED approaches were most commonly used, whereby personalisation in Figure 1 was the most common AIED application.

Figure 1 Zawacki-Richter et al.'s (2019) original AIED typology and use of AI in Institute X

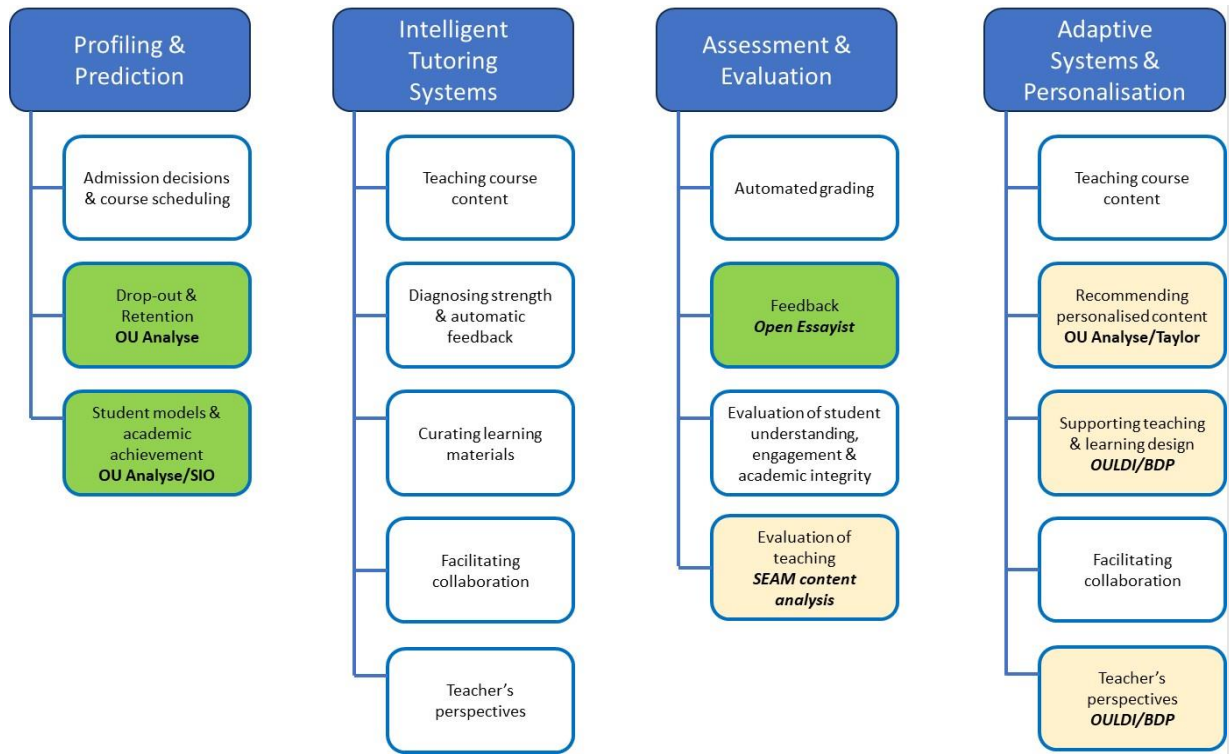

Adapted from [16].

Note: Green colour denotes strong evidence of impact of AI use in Institute X. Yellow indicates some emerging evidence of AI use in Institute X. White colour indicates limited or no initiatives at the moment.

While most AIED studies focus on the design and application of AI in blended and face-to-face settings [8, 16], according to Bozkurt and Sharma [33], "generative AI has the potential to transform distance education and online learning in many ways, including reimagining the roles of educators and universities". Beyond the affordances already identified by Zawacki-Richter, Marín, Bond and Gouverneur [16] of personalisation, intelligent tutoring, and automatic grading, Bozkurt and Sharma [33] argue that in particular distance learning institutions could use Gen AI for content creation, thereby providing more variety and reducing production costs as well as personalised career advice. Furthermore, in a systematic literature review of 26 studies on AI, robotics and blockchain Chaka [5] indicated that distance learning institutions "need to utilize AI-powered chatbots to support students with a view to reducing the ever-present social isolation in online learning environments".

Nonetheless, several challenges for distance learning institutions were identified by Bozkurt and Sharma [33] when implementing AI, including bias in data and algorithms. In particular, given the often wide and diverse student profiles of students at distance learning institutions, relative to "traditional" universities, particular groups of learners might be disadvantaged or inappropriately labelled by AI [28, 34]. Furthermore, there could be an overreliance on AI by students, which might negatively impact critical thinking and independent learning, as well as concerns around data privacy and security. Finally, under the term singularity Bozkurt and Sharma [33] indicated that AI could become too powerful that it raises questions about the role of humans in the learning process, a concern which is also raised by Bommasani, Hudson, Adeli, Altman, Arora, von Arx, Bernstein, Bohg, Bosselut and Brunskill [35] and Hamilton, Wiliam and Hattie [9].

**AIED development and usage in Institute X**

Institute X (blinded for peer review) is a distance-learning institution established in [removed for blind review] in Europe. Institute X's mission is to make education accessible to everyone regardless of their background. For most undergraduate modules, no formal qualifications are required. All modules are delivered in an online or in a blended format. With around 200K students, Institute X has developed a range of AI and analytics approaches to help support teaching and learning. As indicated in Figure 1, there is extensive experience in terms of profiling and prediction, assessment and evaluation, and some emerging evidence in adaptive systems and personalisation. For example, since 2013 Institute X has built and extended a predictive learning analytics system called OU Analyse [17-19, 36], which provides risk-profiles of students to educators based upon four machine learning approaches to educators. 35+ publications have focussed on the reliability, validity, and use of this AI tool by educators, and is one of the few large-scale implementations of learning analytics across an institution [37]. In addition, a Student Information Office (SIO) system has been developed that provides broader risk-profiles of learners [38].

In terms of assessment and evaluation, various systems have been developed over the years using AI. For example, since 2014 Open Essayist was developed and extended to provide automated, interactive feedback system and an acceptable level of support for students as they write essays for summative assessment. [23, 39] Furthermore, several tools on critical reflection and automatic content analysis of student evaluations have been tested [40, 41].

In terms of adaptive systems and personalisation, together with Microsoft, the AI app Taylor was developed to provide students with accessibility needs an easier user interface [28] to disclose disability needs, and to provide appropriate guidance and suggestions for follow-up support. Furthermore, based upon OU Analyse several pilots have been developed to provide personalised recommendations to students [42]. Furthermore, a range of tools have been developed or are under development to give educators AI advice in terms of learning design such as the Balanced Design Planning tool [43, 44]. In other words, while there is some substantial evidence of AI adoption within Institute X, most of these adoptions are focussed on managing data for educators, while limited applications to scale have been provided directly to students.

**Research questions**

With the recent advancements of Gen AI, it is essential that higher education institutions and distance learning providers in particular keep track of the affordances and limitations of AI, and test and evaluate potential AI solutions for and with their students. In this mixed method study, we used the concept of Voice of the Customer (VoC) [15, 45] to explore what distance learning students think about an AIDA that may be able to address some of their teaching, learning, and support needs. Therefore, the following main research question was formulated:

1) What services would distance learning students expect from an Artificial Intelligence Digital Assistant (AIDA), and what are their potential concerns of such an AIDA?

## Methods

### Setting and participants

This explorative two-stage sequential mixed methods study was conducted at Institute X. In the first stage following a demonstration of a possible AIDA distance learning students were interviewed about their expectations of using such AIDA (see next section). In the second stage, students completed an online quantitative survey constructed based upon interview data from the first stage. In this sequential process, both during the interviews and the follow-up surveys students were able to share their perspectives, and compare and contrast their perspectives with those from their peers.

A random sample of 400 students studying at a range of levels and disciplines were extracted from the institutional database. A personalised email was sent to these students on 13 December 2023, followed by two personalised email reminders on 4 January 2024 and 12 January 2024. Potential participants could select one out of twelve possible timeslots to join an interview. These time slots were selected at different times during the week, from morning to evening as many students at Institute X have work and/or family/caring commitments. In total 16 participants selected a time slot, 11 completed the informed consent form, and 10 participants joined one of the online interviews. One possible reason for the relatively small sample was that the study was conducted during the busy festive December and January months.

As illustrated in Table 1, ten students (six women and four men) participated in the interviews, representing various disciplines: five in STEM, two in business, two in education, and one in arts/social sciences. Most were working while studying at Institute X, with diverse educational backgrounds. Notably, many had professional experience and prior study at Institute X, offering potential deeper AI insights than younger students typically included in AI studies. Four were new to Institute X, while six had already successfully completed courses there.

Table 1 Participants of the study (ordered by discipline and age)

| P | Gender | Age | Discipline | New or continuing | Level of study | #Completed courses | Occupational status | Previous education |
|---|---|---|---|---|---|---|---|---|
| I03 | Male | 30-39 | Arts/Social Sciences | Cont. | 2 | 3 | Full time employed | 3 A Lev |
| I07 | Female | 40-49 | Business | Cont. | 2 | 3 | | 3 A Lev |
| I10 | Female | 50-59 | Business | Cont. | PG | 10 | Unable to work | 5 PG Qual |
| I09 | Female | 40-49 | Education | Cont. | 3 | 4 | Full time employed | 4 HE Qual |
| I05 | Female | 50-59 | Education | Cont. | 3 | 3 | Part-time | 2 Less than A |
| I06 | Female | 20-29 | STEM | New | 1 | 0 | Unemployed | 3 A Lev |
| I01 | Male | 30-39 | STEM | New | 1 | 0 | Full time employed | 4 HE Qual |
| I04 | Male | 30-39 | STEM | New | PG | 0 | Full time employed | 4 HE Qual |
| I08 | Male | 30-39 | STEM | New | 1 | 0 | Full time employed | 4 HE Qual |
| I02 | Female | 60-69 | STEM | Cont. | 1 | 8 | Retired | 5 PG Qual |

**Instruments**

*Online interviews*

In the first stage of this sequential design, we undertook six online interviews (3 with 2 participants each, 4 one-to-one). Interviews lasted on average 54 minutes (range 45m17 to 1h16m58). After welcoming participants and reminding them of the overall purpose of the study by author BR, a visual artefact of a five-minute recording of a possible AIDA was screenshared. In this recording, Author JD illustrated three potential examples of AIDA that Institute X might make available in the near future (i.e., automatic feedback on a set of open assessment quizzes of a particular unit in a module, AI generated flash cards for revision, and finally a digital avatar responding in real-time to student academic questions). The primary reason for sharing this visual artefact was to allow participants to activate any prior knowledge or experience with AI, and if participants had no experience with AI to prompt some initial thoughts and ideas of how AI might support their studies.

Subsequently, the following interview questions were raised in a semi-structured manner: 1) what feelings and/or thoughts do you have as a student in terms of such an AI Digital Assistant (AIDA)?; 2) what services would you expect from such an AIDA?; 3) what would you worry about when using an AIDA. After each discussion of a respective question, the evolving insights from other participants who had participated already in the interviews were shared in the form of text bullet points by the interviewer in a PowerPoint screenshare. At the first interview we shared the perspectives from the research team, and adjusted the information afterwards based upon the input from students. This allowed participants to sense check their answers, and further discuss any agreements or disagreements for each question. In other words, even if a participant was the only one joining an interview, they could compare and contrast their thoughts and ideas with other participants after they shared their initial thoughts.

*Online surveys*

In the second stage of this sequential design, building on the work of [28] and the responses from the participants, a follow-up survey was shared with the 10 participants on 19 January 2024 after the final interview was completed. This survey aimed to gather feedback from participants about the main results from the interviews, the key themes identified, and whether (or not) these resonated with them. Students received a personalised invitation to fill in this survey via email. It consisted of 26 closed Likert response questions constructed based upon the responses from participants during the interviews (1 = Totally Disagree, 5 = Totally Agree), including four based on expected services of AIDA (i.e., support for academic tasks; real-time assistance and query resolution, personalisation and accessibility, and emotional and social Support; Cronbach α = .686), and five broad concerns about using AIDA in teaching and learning (data privacy and use; academic integrity; operational challenges; ethical and social implications; and future of education; Cronbach α = .687). Five open questions to gather further qualitative feedback and suggestions were included. In total 8 out of 10 participants responded to this follow-up survey in January 2024.

**Procedure**

All interviews were conducted by BR, an experienced mixed methods researcher. Interviews were conducted online using MS Teams. Audio was automatically transcribed (with explicit permission from participants). Transcripts were then checked, and cleaned by BR. Extensive information about the study was shared with participants prior to the start of the interview, and participants provided informed signed consent prior to the interview.

All transcripts were subsequently anonymised and were uploaded per interview question in ChatGPT 4 on 17-18 January 2024 by BR. In line with recommendations of [11], [46] and [47], we initially used emergent theme analysis within ChatGPT4 to generate separate analyses for each of the interview questions, followed by a range of prompts to position these themes with respective interviewees, as well as their positive, negative, or neutral sentiments. These themes afterwards were sense-checked by three authors (BR, CH, FT) and overall had good face-validity, and resonated well with the notes and analysed transcripts. Themes were subsequently merged and grouped together where feasible by the three authors. These data were used to develop the online follow-up survey, whereby the participants could check and validate whether the emergent themes resonated with them (or not). SPSS 27 was used for the online survey results. This research received Human Ethics Research Approval (HREC/XXX). Participants were free to participate and withdraw their written consent at any time. No consent was withdrawn.

# Results

## Results from the interview: expected services from AIDA according to students

In stage 1 after watching the visual artefact of a potential AIDA for five minutes the initial feelings and thoughts of the ten participants were explored in the first interview question. In general, most participants were fairly surprised and intrigued that such an AIDA would work, in particular as towards the end of the screenshare JD indicated that the digital avatar was not a recording of him but an AIDA. Nine out of ten participants indicated that an AIDA would be useful for their studies, in different ways as indicated below, while one participant (I10) could see it being using for students with disability or support needs, but not for her own studies.

Four main emergent themes were identified in terms of expected services that students would find useful for their teaching and learning. In terms of the most mentioned service by participants, which we label as *real-time assistance and query resolution*, most participants expressed the support for the idea of having 24/7 support from an AIDA for academic queries and guidance. For example, I09 indicated that such an AIDA could provide 24/7 real-time support when she needs it.

> I think 24/7 support would be good because we all work it very, very different times. And what I would be expecting is related to the course material. So, if I had a question on "where would I find information on the red bus", [AI] would give me links to where I could find this information, so that I'm not going at the tutor all the time and saying where can I find this? (I09, 17:36)

Also I01 indicated that immediate feedback would be helpful to find appropriate resources, and get more course relevant information.

> One benefit is you have immediate questions and response. I would see this go a little bit further. Like when you ask a question to search something from the university, it'll be helpful if within the results, instead of just having a plain explanation, to point out where I can search more, or where I can find more details about this question. (I01, 11:08)

Similarly, I08 indicated that beyond the quizzes illustrated in the visual artefact of AIDA a chatbot-like function to explore real-time academic queries might be useful.

> I think the ability to have a chat bot that you can prompt while you're revising is very good and cause quite often when I revise or I'm studying for an [assessment] I get stuck on the topic and maybe I don't have time to speak to my tutor unfortunately. Or I can't find the

> answer I'm looking for in one of the forums to be able to have a chat bot that I can prompt and discuss with for revision purposes. (I08, 16:29)

At the same time, I09 indicated that there needed to be a balance between guidance and directly giving the "correct" answer.

> It's just making sure that we've got a grasp of what we're learning. It's a difficult balance to get that kind of support without specifically giving out answers, because I know the AI won't tell you the answer to the question, it is pointing to the right direction as what I'm looking for. Not giving me the answers, just putting you in the right direction, that sometimes that's all you need. (I09, 18:36)

In terms of the second most expected service, which we label as *support for academic tasks*, most participants were positive that AIDA could support them with particular academic activities, such as providing summaries of main points from study materials, providing feedback on learning activities, language support, and offering suggestions for (additional) study materials.

> It should be able to cover the entire topic of the module, obviously. To have knowledge about the resources, tutorials, the library, so the student will be able to interact easily and find the information that are needed quickly... So when I speak with the [AI] if it would be possible to "OK give me the answer that I asked the question", at the same time point me where I can read more. (I01, 21:10)

Beyond finding resources on academic progress, I03 would like the AIDA to provide some academic language support.

> The very first basic service that I would like to have is finding sources, the other one, and to me this is one of my main usages of AI, I'm not a native and English speaker, writing documentation, even for my work, sometimes looking at essays, I'm really not 100% confident about my grammar. I would like the AI to tell me what I'm doing wrong, and how can I do better (I03, 16:55).

In terms of support for academic tasks around half of the participants were wondering whether such AIDA could support multiple languages. As three participants (I01, I03 I06) were international students and non-native English speakers, they found it useful having an AIDA that would work in different languages. Preliminary evidence shows that Gen AI approaches could be useful for language acquisition of non-native English speakers [30]. At the same time, some like IO9, who is a native

English speaker but from a particular region in the UK, was worried about whether or not AI would pick up her respective accent.

> I actually I like the idea of it. I like the idea of having like a 24-hour support system. One of the issues that would come up for me would be my accent on the speech system. No matter where I'm, it's the accent, always, always get issues with that. And I'd really like the recommendation of the AI on activities. (I09, 7:44)

At the same time, I03 would see the AIDA as an assistant, not a replacement of a human tutor.

> I would love this to be like an assistant, not the teacher. Sometimes I have some questions that I would like to verify with the teacher. I understand the teacher sometimes is busy, they cannot answer a question right away. I do not want the AI to replace the tutor but be an assistant and sometimes help me to refine information but do not block me to talk with the tutor. (I03, 18:22)

Other academic support included the opportunity to quiz oneself, as for example indicated by I05, I06 and I07:

> I'm a big fan of the quiz. I love that, especially if you're not in a class full of people that might laugh at you when you get all the questions wrong. It's a really good way of reminding you a) what you need to know, like the questions you need to be asked cause that's one of the things I forget in my studies is what I need to be actually answering, and b) reminding myself how much I've actually learnt. (I05, 13:32)

> Sometimes it is tricky for me to revise, like I don't know exactly what's the best way how to do it, like flash cards, or I don't know, talking with other people, but it's not as interactive, just doing it by myself, I feel it's a bit tricky sometimes. (I06, 9:09)

> I actually wish that existed right now. I'm really struggling with my coursework and I don't know where I'm going wrong and they're all seems to be almost loops that I miss. (I07, 7,59)

Finally, some indicated that summarising main points of research papers and Institute X learning materials could be useful.

> One area which I think is super useful for us for summarizing research. You might spend a couple of hours reading a 30-page research paper and to get all the points from, I think if AI teaching assistant would have the ability to summarize a research paper for you and take the key points out of it, I think something like that would be very useful. (I08, 15:26)

The third expected service of *personalisation and accessibility* referred to participants expressing a desire for AI to be personalized to their own individual needs and learning approach, including those with accessibility needs. IO6, for example, indicated that students might have different study motivations, and for her it would be useful to get support of what you know and what knowledge gaps might still remain.

> Finding out gaps in my study because technically you could fly through Institute X, missing huge chunks of information, but you just have to say the right stuff. If you're just in it for the degree, good for you, but if you are in it to learn and gain knowledge and expand your understanding of your certain subjects, you need to have someone to tell you what you don't know... I'm just annoyed that [AIDA] doesn't exist right now." (IO6, 13:24)

Others indicated that AI could provide personalised support for students in need.

> In terms of student support, such as disability support and stuff like that, and thinking about this being available 24/7, I could imagine AI might be quite handy for some students. That's stressing out in the middle of the night, and they can't find stuff about the deadlines and submissions, or how you meant to format a document and stuff like that, or who do they contact if a student breaks down. (I10, 19:13)

In terms of the fourth and final service of *emotional and social support*, a mixed perspective emerged about the potential of AI to offer emotional support or motivation. Some like I05, I06 and I07 thought that such service from an AIDA could be helpful in some circumstances. For example, IO7 did not mention emotional and social support initially, but when she was prompted by the interviewer based upon what some other students indicated, this did resonate with her as she lived remotely:

> I think [emotional and social support] would be great. I know there are a lot of people who study with Institute X because they are very anxious in social situations, or they just can't be around huge groups of people, so [Institute X] is the choice for them. Or maybe they just live in a very kind of rural area and the cost of going to a "bricks-and-mortars" university like having to travel on the train everyday, or having to move house, or rent a house somewhere. It's not worth it so than you're pretty isolated. (I07, 15:54)

Also, I05 and I06 mentioned that an AIDA could help with social isolation. In particular as some social media groups that are run by students from Institute X are not necessarily supportive.

> I think both I06 and I have mentioned the sort of isolation of distance learning. If I knew it was like an AI at the other end I think it would help me just to have somebody to talk to sometimes. You know not everybody likes WhatsApp groups and Facebook, and certainly I've heard about people being bullied on Facebook. (I05, 24:30)

At the same time, some argued that there is a limit in terms of what AI might do to provide emotional and social support.

> I think there has to be a limit of support socially or emotionally [be]cause if someone really needs like professional help I don't think AI is capable, right now, to give that, and it's good to actually talk and to be with other people. (I06, 20:31)

> One thing that concerns me, it is this development of social and emotional skills because this can sound to me like AI is replacing a therapist. I am a neurodiverse person so I am ADHD. What works perfectly well for the majority of neurotypical people simply doesn't work for me. When you automatize this through an AI you can lose these nuances of like emotional and social skills. (I03, 22:20)

**Results from the interview: potential concerns of AIDA according to students**

In terms of potential concerns of making AIDA available to students five themes emerged from the interviews. Most participants were concerned about the *ethical and social implications* of implementing AIDA, that is the potential impact of AI on learning processes, and the necessity of balancing technological advancement with human-centric educational practices, and keeping the human (student/tutor/academic staff) in the process and being able to talk to a human.

Participants expressed a desire for the AI to assist, not replace, human interaction in distance education, including an appropriate balance between AIDA and human tutoring [9, 28]. Within Institute X so-called associate lecturers (AL) are assigned to students, whereby each AL supports around 20 students per module, and often students develop a strong relationship to their tutor at a distance. For example, I01 and IO3 indicated the negative potentials of AIDA of replacing human tutors:

> The tutor should be always present at some point, like if I need to contact for a specific reason I should be able to do that. (I01, 15:32)

> My biggest concern with all the usage of the AI and also how much we are looking to the AI as a tool that can really help us not as a tool that is just replacing humans… Talking with a computer but believing I am talking with human, I think it is very concerning. (IO3: 10:28)

I10 was worried that the training of such models might introduce huge ethical concerns, in particular in terms of which world views are shared, generated, and spread within AI.

> With each assignment it will develop, and work out which is the best assignment, so in terms of Equality, Diversity, and Inclusion (EDI) I'm thinking: "What is it learning? The Western model of the mind? The male model of the mind? These universities are very discriminatory!" So that certain world views take even more control of what knowledge is desirable, and as a university, particularly somewhere like Institute X , they should be, there should be really quite concerned about that sort of stuff. There are things that once it gets institutionalized and become dogma. So it's got to be right! That is a scary world! (I10, 40:46)

Others like I05 and I06 had a different perspective and were mostly positive about the affordances of AIDA, and having 24/7 access to a personalised AI tutor, but they did express concerns that it might lead to an overreliance of AI rather than human support.

> I wouldn't have much to worry about, the only thing like we would just saying about the people relying on it, and to the extent of ignoring live tutorials and things. I'm a big sci-fi fan and I just can't wait until AI rules the world, so I have no worries about it all… And you know with like gambling sites, they have reminders that "you've been gambling for all day, you should probably stop!". There's maybe like hints, you know, saying "Ohh you know you've been talking to me for a while now. Would you be interested in contacting your tutor? (I05, 23:10)

In terms of the second most explored concern by participants, *data privacy and data use*, in line with wider literature, participants were worried about their privacy [1, 11, 48], the engrained potential biases in AI [8, 49, 50], and its potential impact on diverse student populations [34, 51, 52]. Some participants indicated that they would be worried their data being used for training an AIDA system, and what would happen with their data. For example, IO3 works in the gaming industry and was worried how creative outcomes of students might be compromised.

> Another worry, it is basically privacy of usage of my data, and this is my biggest worry with like ChatGPT, not only ChatGPT, but Midjourney for instance. I talk with artists every day and sometimes they are quite upset with, like, "OK, they're just stealing this artist". I think it is interesting if Institute X might have … my information without my consent to train the AI, I mean I would be OK most of times to use it, but I think it's important to be clear to me, like: "hey [I03], we want to use these inputs to train the AI. Are you OK with that or not? (I03, 27:43)

Similarly, two participants expressed concerns about intellectual property of AI-generated content. For example, I08 expressed concerns about who owns the data submitted to an AIDA.

> If Institute X was to have a teaching assistant where I was plugging in my work and getting an output, like who owns that? And is that used to train the model? I see companies now coming out with "you can pay to not have that data used to train the model and vice versa", but you know, is that a good thing or is that a bad thing?

As already reported widely in the media and literature, Gen AI systems could start to hallucinate and generate incorrect or biased answers [9, 11]. In total two participants mentioned this unprompted (I03, I08).

> The level of hallucinations that an AI tool creates. I was reading what ChatGPT was saying, and it was like, OK, this is clearly wrong! If I was not trained on this, I would believe you're right, and this is really concerning because it can spread wrong knowledge (IO3, 13:01)

The third theme related to *operational challenges* of implementing such an AIDA, as AI might inadvertently affect learning processes and student interactions, and making sure that the AI tools are accurate and reliable. Some participants like I03 and I08 expressed a need to be very clear when information provided to students came from AI:

> Not knowing that I'm talking with the robot. It is really, really important. I will feel kind of betrayed if I believe I'm talking with a human, and at the end I realise it is a robot, so I think if you are playing with the teacher avatar like you showed at the beginning [of the interview], it is really important for the avatar to say: "hey, I'm just an avatar. I'm not a real teacher!". (I03, 27:07)

Linked to this point, given the wide range of disciplines and topics supported in Institute X, I08 was worried about how AIDA would be able to cover the wide range of disciplines.

> I think that's a risk as well, with the broad amount of subjects and teachers at the Institute X, to make sure it's correct 100% of the time I imagine it would be a big challenge and would be worrying [the AIDA research team]. (I08, 26:23)

A fourth theme centred around *academic integrity*, and potential misuse of AI by students for completing assignments and potential plagiarism. For example, I07 indicate that the AI systems could be trained with incorrect data from students.

> But what if your work is rubbish? Would that not like battle [for] influence? The AI then would go teaching incorrect information to other people? (I07, 25:55)

I08 indicated that it might lead some students to misuse the system in order to get the right answers on assessments rather than to learn new knowledge and skills.

> It would need to be entered into the way that you have prompted to give you answers to things. Because I think you know when you do degree, you're trying to learn. And I think maybe the more mature you are as a student, the more the less prone you probably are to try and prompt an AI system to give you the answers to your assessments. But you know there is a risk in this as well. (I08, 9:46)

IO9 expressed that she was worried about plagiarism when using AI, and not being able to learn.

> I don't play with them because I'm very conscious of plagiarism and all, and I know they're designed to do that. But the work supposed to be mine and I'm not learning if it's not my work. (I09, 14:11)

Finally, some participants were concerned about the *future of education*, and how AI integration might change the nature of education and assessments, necessitating a shift in teaching methods and learning expectations. For example, I08 indicated some potential long-term impacts of AI that might influence how companies might look for economic efficiencies by replacing humans with AI, and what the impact might be on his children.

> I'm someone who worked as a financial controller for many years, and you start to automatically look for financial efficiencies. Could you replace people with essentially avatars and chat bots? Which is scary! [A] lot of people would need retraining and a lot of different things, which does worry me. I think about my children and what kind of world they need to train for when they are older. (I08, 24:44)

I10 was specifically worried about the impact of AI on critical thinking. Some of this was also mentioned in different words by others like I08, but I10 mentioned this point six times during the interview.

> One of the problems I see with all students is that they lack criticality. So I don't think it's doing them any favours to make them become more insular and more focused [when supported with AI]. All they usually interested in is how to write a good assignment to get a top mark, so they're not even interested necessarily in the subject. Now we're in a very complex global environment, so this isn't exactly equipping them with the right skills."

**Online survey outcomes**

The second stage of our data collection included a follow-up survey two weeks after the interviews, whereby in total 8 out of 10 participants responded in January 2024. By sharing four main services that students might expect from AIDA and five concerns discussed during the interviews participants had the opportunity to reflect on their own perspectives and those from other participants, as well as add any new elements that were not discussed during their interview. At the start of the survey, participants were asked whether they would like an AIDA for their studies as demonstrated by JD. 5 out of 8 participants agreed with this statement (3 indicated totally agree), two were neutral, and one totally disagreed.

In terms of the types of services that participants would like AIDA to have, as indicated in Table 2 the most popular was *real-time assistance and query resolution*, followed by *personalisation and accessibility*, and *support for academic tasks*. While there was substantial support by most participants for these three services, only 4 out of 8 participants agreed that having *emotional and social support* should be a feature of AIDA, 2 were neutral, and 2 disagreed.

Table 2 Four services that students expect an AI digital assistant to have for teaching and learning

| | Mean | SD | % agree |
|---|---|---|---|
| *Real-time Assistance and Query Resolution*: Having 24/7 support from AI digital assistant for academic queries and guidance | 4.25 | 1.04 | 87 |
| *Personalisation and Accessibility*: AIDA to be personalised to individual needs and learning approaches, including those with disabilities or specific learning requirements. | 4.13 | 1.48 | 75 |
| *Support for Academic Tasks*: AI assisting with academic activities like summarising key points from study materials, providing feedback on assignments, helping with grammar and writing, and offering study resources. | 3.88 | 1.12 | 62 |
| *Emotional and Social Support*: AIDA provides emotional support or motivation if needed, especially in the context of distance learning or for students with social anxieties. | 3.25 | 1.28 | 50 |
| *Real-time Assistance and Query Resolution*: Having 24/7 support from AI digital assistant for academic queries and guidance | 4.25 | 1.04 | 87 |

n = 8, α = 0.686.

In terms of additional qualitative feedback provided by participants, for example I07 indicated that this would be a useful addition to her studies: "I think all the points are very valid and relevant tools to improving students' studies". (I07)

I08 agreed but he was worried about the emotional and social support.

> I think the AI digital assistant has many great use cases, mainly the first three that are listed above. However, I do not feel that an AI digital assistant should be available as an emotional or psychological support, there are too many undefined risks surrounding this. Including human attachment to a non-physiological form and the risk of a student suffering from emotional distress not meeting with other humans. (I08)

I05 also indicated that she would see value for all four AI features, but that she would not need them for her studies as she is already an experienced learner.

> I have chosen 'totally agree' for the areas I'd most likely use and 'agree' for those I think would be great for others, but I personally don't feel I'd need as much. (IO5)

In contrast, I10 remained sceptical whether such an AIDA would be useful for her studies.

> I support life-long learning and using the human brain. This just sounds like another form of social control. (I10)

In terms of the potential concerns of using AI in education, as indicated in Table 3 participants were most worried about *Academic Integrity*, *Operational challenges*, and *Ethical and social implications* of AI. 7 out of 8 were worried around potential misuse of AI for completing assignments and potential plagiarism, and similarly participants were worried about the operational issues of using AI (e.g., how

AI might inadvertently affect student interactions and learning outcomes, how accurate and reliable the tool might be). Furthermore, 6 out of 8 were worried that AI might have an impact on learning processes and the role of human contact in teaching and learning.

Table 3 Concerns about use of AI in education

|  | Mean | SD | % agree |
|---|---|---|---|
| *Academic Integrity*: potential misuse of AI for completing assignments, potential plagiarism, and academic integrity | 4.38 | 0.74 | 87 |
| *Operational challenges*: how AI might inadvertently affect learning outcomes and student interaction, and the importance of ensuring that AI tools are accurate and reliable | 4.25 | 0.71 | 87 |
| *Ethical and social implications*: the potential impact of AI on learning processes and the necessity of balancing technological advancement with human-centric educational practices, keeping the human (student/tutor/academic staff) in the process, being able to talk to a human | 4.13 | 0.84 | 75 |
| *Data privacy and use*: how student data and inputs to the AI are used, stored, and potentially shared, emphasising the need for transparency and consent | 4.13 | 0.84 | 75 |
| *Future of education*: how AI integration might change the nature of education and assessments, necessitating a shift in teaching methods and learning expectations | 4.00 | 0.76 | 75 |

n = 8, α = 0.687.

This was followed by concerns about *data privacy and use*, and the *future of education*. For example, I05, who was a self-proclaimed enthusiast for tech, indicated that there might be a potential risk of some students becoming too reliant on such AI technologies.

> I feel fairly neutral about most of these although I do think there could be a potential risk of people relying too heavily on AI for both emotional and academic support. I think the information given needs to be very carefully looked at, students do need to work things out for themselves after all. (I05).

Similarly, I07 indicated that 'these are important topics. Its understandable people will be worried about their information and how it will be stored'.

## Discussion

With the rapid evolution of AI applications in the last five years, in particular in the field of so-called Generative AI (Gen AI), many higher educational institutions and distance learning institutions in particular are exploring how to effectively support their students [1, 3, 27, 33]. Specifically with the Bloom dream of providing one-to-one personalised learning opportunities to increase student attainment using Large Language Models like ChatGPT, there are a lot of expectations around AI providing appropriate personalised learning at scale in the near future [8, 12].

In this mixed methods two-stage sequential study, using a Voice of the Customer approach [15, 45] we demonstrated a possible AI digital assistant (AIDA) to ten distance learners at a large distance learning provider in Europe with detailed interviews, and subsequently a follow-up online survey to validate their responses. We explored what distance learners considered to be some of the expected services of using AIDA for their studies, and whether they had any concerns using AIDA for their studies.

Four expected services were identified that distance learning students would find useful for their teaching and learning: 1) real-time assistance and query resolution; 2) support for academic tasks; 3) personalisation and accessibility; 4) emotional and social support. Most participants were expecting to get 24/7 support from an AIDA for academic queries and guidance, in line with Bozkurt and Sharma [33]. In particular participants were expecting that AIDA would provide immediate feedback on their knowledge and skills [8, 16], and find appropriate resources and course relevant information where needed [12, 24]. At the same time, several participants indicated that there needed to be a balance between AIDA providing guidance and directly giving the "correct" answer [53]. In terms of support for academic tasks, most participants were positive that AIDA could support providing summaries of main points from study materials, providing feedback on learning activities, and language support. As a third service, participants expressing a desire for AI to be personalized to their own individual needs and learning approaches, including those with accessibility needs [27, 28].

Finally, there were mixed expectations in terms of emotional and social support, whereby some students absolutely would embrace such service, while other were completely against Institute X considering this at all. In particular, in distance learning contexts where there is often social isolation due to the distance learning provision [5], as well as a large group of students specifically choosing to study at their university because of their accessibility needs [28], this service in particular would require both more attention as well as careful consideration whether (or not) this might be a desirable feature.

In terms of linking these expected services with Zawacki-Richter, Marín, Bond and Gouverneur [16]'s original AIED typology, several of these services aligned well with specific elements within the typologies of intelligent tutor systems, assessment and evaluation, and adaptive systems and personalisation. At the same time, the way that students talked about the types of services they were expecting seemed to span activities across the typologies, rather than fit in one or another typology. This might support the notion of [8] who raised concerns that most of the AI research focussed on undergraduate STEM and Computer Science students studying at on-campus universities. By using a mixed methods approach that included students' perceptions of AI in a distance learning context, we provided some new narratives of what students might be expecting in terms of AIDA, and what their concerns are for their teaching and learning.

Regarding concerns, most participants expressed worries about the ethical and social impacts of AI in education, stressing that its role should enhance, not substitute, human interaction. They highlighted the risk of AI promoting biased perspectives and the issue of overreliance on technology, which could diminish the importance of human support and interaction [8, 49, 50]. Concerns about the dissemination of incorrect information and the potential adverse effects on diverse groups were also highlighted.[34, 51, 52]. Bias and discrimination are key issues in debates about the use of AI in education especially the reproduction of existing power structures and inequalities in algorithms [9, 28, 54]. Additionally, participants expressed concerns around data privacy, how personal information would be used in AI systems, and issues of data and intellectual property ownership [1, 11, 48] . This is particularly relevant for distance learning contexts as students mostly learn online, and therefore their activities and behaviours online are more traceable relative to on-campus students. Following these points, operational challenges and academic integrity concerns seamlessly emerged, emphasizing the critical need for AI's reliability and highlighting the risks associated with its misuse in education. Finally, participants also discussed the impact of AI on the future of education, particularly focusing on the potential impact on teaching methods and critical thinking.

A unique contribution of this study was the voice of the customer (VoC) approach, whereby we have focussed on distance students from a range of ages, academic progression, disciplines, experience, and levels of study. Another unique contribution was the fact that most participants had already extensive professional experience, thereby providing more complex and perhaps comprehensive perspectives beyond "just getting a degree". In particular, the narratives around the future of education and whether or not higher educational institutions should be critical towards AI adoption require substantial attention, in line with recommendations by Hamilton, Wiliam and Hattie [9].

**Limitations and future research**

A main limitation was the relatively small sample of this study. However, we reached saturation in terms of the themes that emerged after eight participants, and given the validation of the subsequent results in stage two by the participants there was consensus amongst most participants in terms of the expected services of AIDA and its potential concerns. In future research, in line with design-based research principles we would like to explore how initial prototypes of AIDA are perceived by both students, educators, and managers. This would help to develop inclusive and transparent mechanisms and policies with all stakeholders involved to make sure that a future AIDA approach is appropriate for the diverse range of students at Institute X.

## Acknowledgements

We are grateful for the support from all those students who participated in this research.